# Criticality Aware Multiprocessor

by
Sandeep Navada and Anil Krishna

## 1. Overview

Typically, a memory request from a processor may need to go through many intermediate interconnect routers, directory node, owner node, etc before it is finally serviced. Current multiprocessors do not give preference to any particular memory request. But certain memory requests are more critical to multiprocessor's performance than other requests [4][5]. Example: memory requests from critical sections, load request feeding into multiple dependent instructions, etc. This knowledge can be used to improve the performance of current multiprocessors by letting the ordering point and the interconnect routers prioritize critical requests over non-critical ones.

## 2. Implementation

There are four main parts to the implementation of this project.
- Building criticality-aware interconnects
- Identifying critical memory requests
- Creating a lock intensive microbenchmark
- Building hypercube interconnect

We have implemented a second set of virtual networks (and the associated message buffers, ports, actions and transitions, as necessary) to carry critical requests and responses. We extended the *MOESI_SMP_directory* protocol to contain these new set of virtual networks, and called it *MOESI_SMP_directory_crit*. We also implanted a per-processor variable which identifies whether the thread running on the processor is inside the critical section (or, more generally, executing critical code). This variable, called *crit*, was implemented as a member of the *CritSecInfo* class; an instance of an object of this class was declared for each processor. Testing that the requests traveling on the critical virtual network are actually given a higher priority required distinguishing critical requests from non-critical ones.

In order to selectively identify some requests as being critical, we use the MAGIC instruction support in Simics. The MAGIC instruction is used to trigger a call-back function called *toggleCrit()*, which is a public method in the corresponding processor's *CritSecInfo* object. This method toggles the *crit* variable. Upon entering a critical section (acquiring a lock) *crit* is set to *true*, and upon leaving a critical section (before releasing a lock), *crit* is set to *false*.

Our microbenchmark contains a shared array of counters which is incremented in the critical section. The microbenchmark was created using pthreads. The magic instruction was inserted just after the lock and just before unlock. Also magic instruction for loading ruby module, clearing the stats and dumping the stats was inserted at appropriate places.

For our evaluation, we wanted to explore hypercube which was not included in ruby. We implemented Hypercube using the source code of 2d Torus.

# 3. Evaluation

## 3.1. Experimental Methodology

Baseline for evaluation is as follows:
Number of Processors: 16/4
Processor Core: In-order
L1 Cache Size: 256 KB
L1 Associativity: 4
L2 Cache Size: 16 MB
L2 Associativity: 4
Block Size: 64B
Memory Size: 512 MB
Coherence Protocol: MOESI SMP directory protocol
Bandwidth: 125
Network Topology: Crossbar/ 2D Torus/ Hypercube
Microbenchmark: Incrementing shared array of counters
Simulator: SIMICS/GEMS[1][2]

Please note that a low bandwidth system (125) is chosen to model a system in which bandwidth available is scarce.

## 3.2. Performance with microbenchmark

The first experiment shows the speedup obtained by Criticality Aware Multiprocessor (CAM) over the baseline multiprocessor.

| Configuration | Cycles(baseline) | Cycles(CAM) | Speedup |
|---|---|---|---|
| 16p.CROSSBAR | 152067271 | 144425566 | 1.052911027 |
| 16p.TORUS_2D | 159040827 | 143800329 | 1.105983749 |
| 16p.HYPERCUBE | 168426883 | 147134017 | 1.144717492 |
| 4p.CROSSBAR | 83345936 | 76637069 | 1.087541 |
| 4p.TORUS_2D | 86058039 | 77021911 | 1.117319 |
| 4p.HYPERCUBE | 86058039 | 77021911 | 1.117319 |

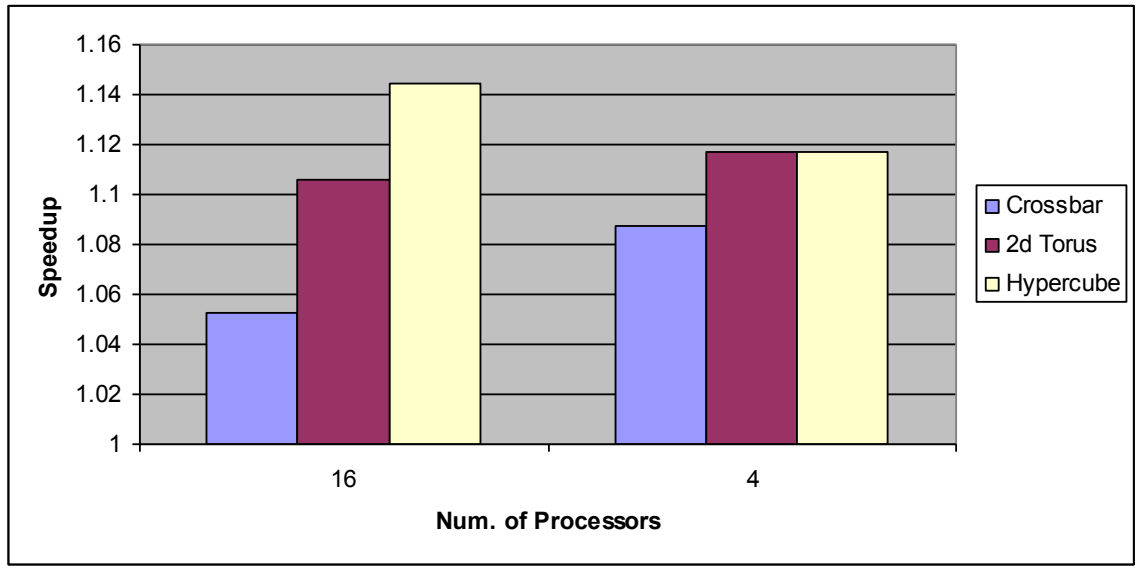

We see that CAM gives a speedup of 5-15% speedup over the baseline multiprocessor. It is interesting to see that the CROSSBAR topology performs better than HYPERCUBE/2D TORUS. The reason behind this is that the switch being modeled is a perfect switch (that is there is no contention modeled at the switch). Hence, for a CROSSBAR, there is a dedicated path between any two processor nodes which is not the case in HYPERCUBE/2D TORUS.

## 3.3. Sensitivity studies:

### 3.3.1. Effect of number of critical requests

We change the number of critical requests by changing the number of shared array counters.

We decrease the number of shared counters to a third to study the effect of number of critical requests on performance.

| Bench.Proc. Topol | Cycles(baseline) | Cycles(CAM) | Speedup |
| --- | --- | --- | --- |
| micro.16p.CROSSBAR | 152067271 | 144425566 | 1.052911027 |
| micro.16p.TORUS_2D | 159040827 | 143800329 | 1.105983749 |
| micro.16p.HYPERCUBE | 168426883 | 147134017 | 1.144717492 |
| micro(1/3).16p.CROSSBAR | 88741356 | 86323484 | 1.028009435 |
| micro(1/3).16p.TORUS_2D | 91746798 | 83569596 | 1.097849007 |
| micro(1/3).16p.HYPERCUBE | 93135640 | 83915377 | 1.109875726 |
| micro.4p.CROSSBAR | 83345936 | 76637069 | 1.087540757 |
| micro.4p.TORUS_2D | 86058039 | 77021911 | 1.117318928 |
| micro.4p.HYPERCUBE | 86058039 | 77021911 | 1.117318928 |
| micro(1/3).4p.CROSSBAR | 62638930 | 57866424 | 1.082475 |
| micro(1/3).4p.TORUS_2D | 65266363 | 61008605 | 1.069789 |
| micro(1/3).4p.HYPERCUBE | 65266363 | 61008605 | 1.069789 |

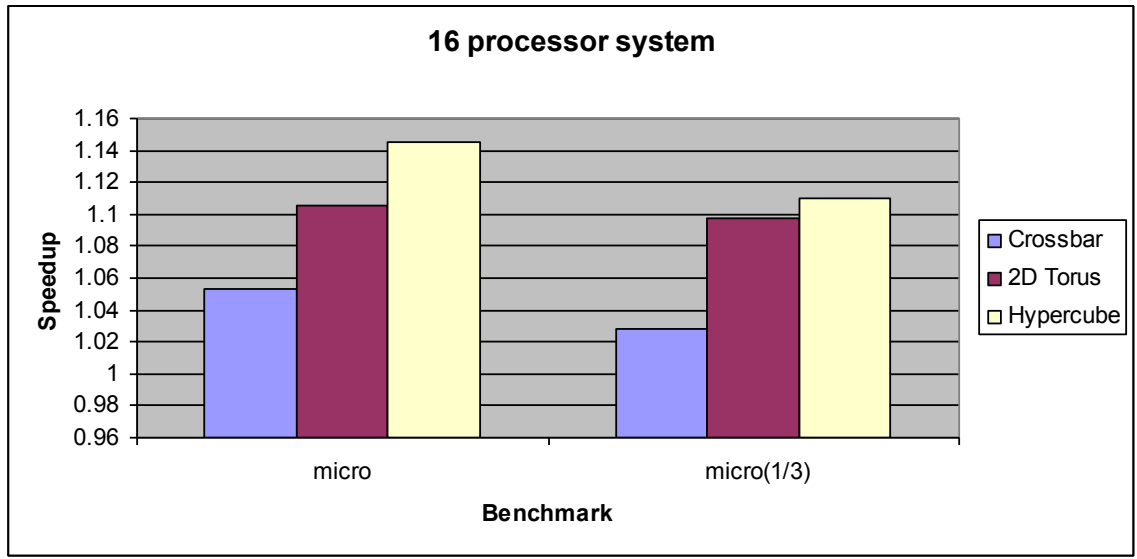

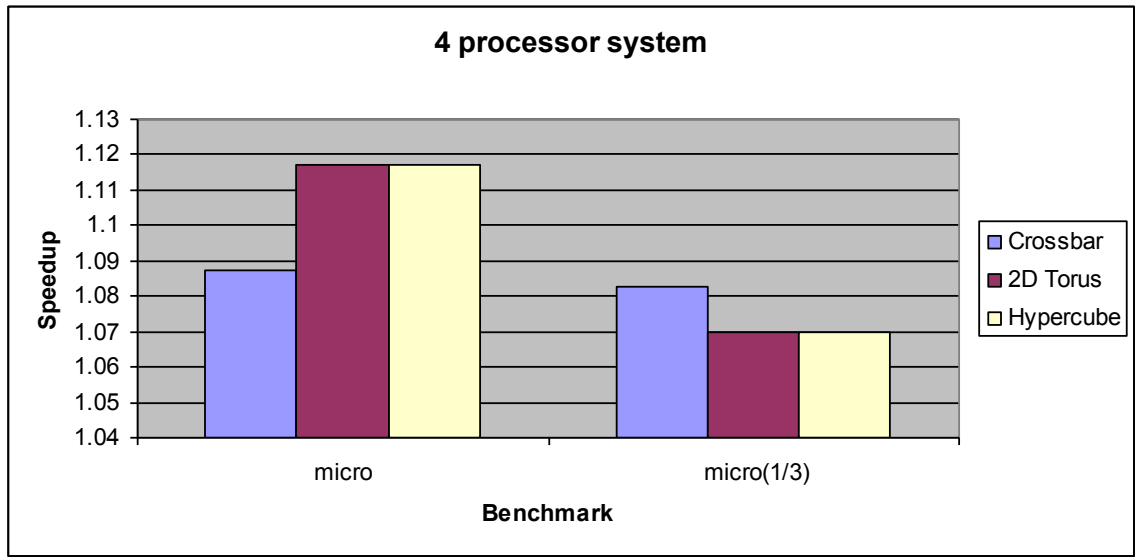

We see that the speedup obtained by CAM in both 16 processor and 4 processor system decreases when the number of shared counters is decreased. When the number of shared counters is decreased, the number of critical request also decreases resulting in lesser performance gain.

To confirm this fact, we instrument counters in the simulator to find the ratio between critical request and non-critical request.

| Bench.Proc | Crit. Req. | Non-crit. Req. | Ratio |
| --- | --- | --- | --- |
| micro.16p | 298038 | 479900 | 0.383113 |
| micro(1/3).16p. | 101396 | 376908 | 0.211991 |
| micro.4p. | 99482 | 434150 | 0.186424 |
| micro(1/3).4p. | 26268 | 213723 | 0.109454 |

The above table confirms that the ratio of critical requests and non-critical requests decreases when the number of shared counters is decreased.

### 3.1.2. Effect of changing the bandwidth

To see the effect of changing the bandwidth, we increase the bandwidth of the system to 250.

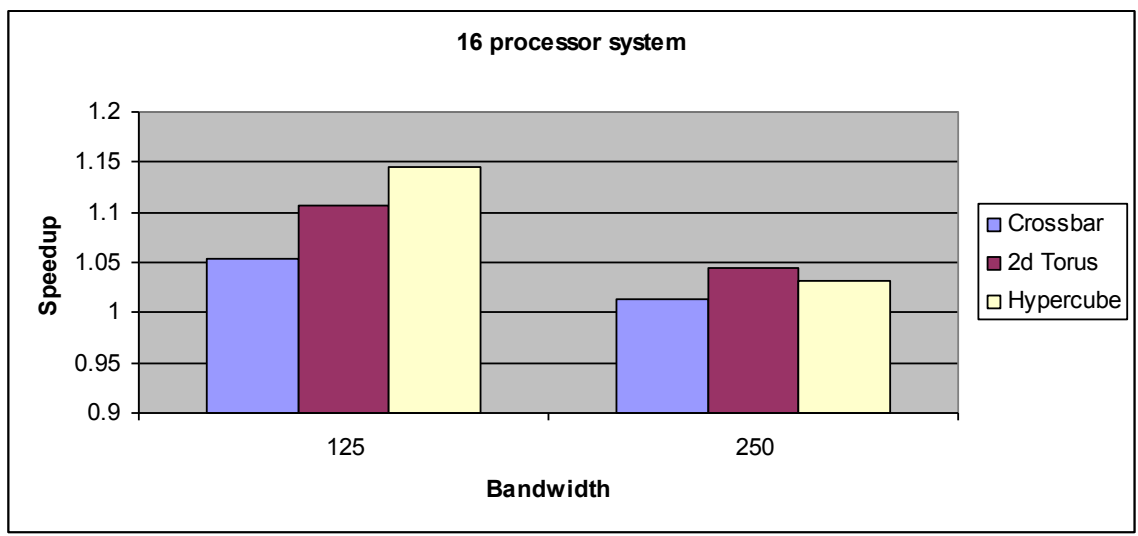

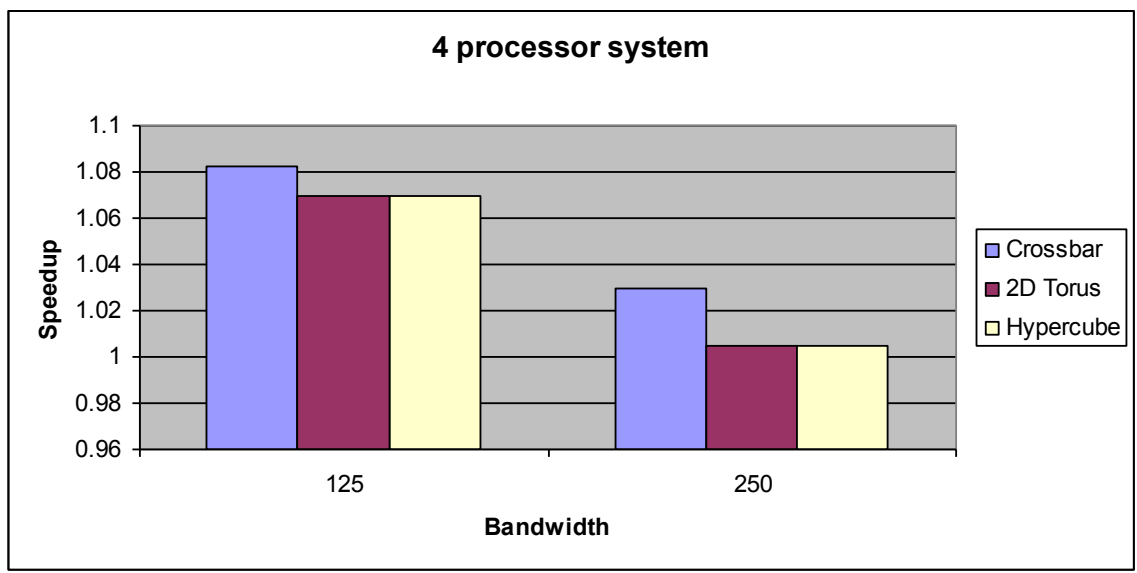

We see that the speedup obtained by CAM over the baseline processor is drastically reduced when the bandwidth is increased from 125 to 250.

To understand this effect, we try to look at the average link utilization and the average contention cycles for 16 processor system. We define contention cycles as the number of cycles in which the there are both critical and non-critical request present at the input buffer of the link. As the switch modeled is perfect, hence CAM will be able to prioritize critical request over non-critical request only at the input buffer of the link (throttle).

| Bench. bandwidth | Avg. link utilization | Avg. contention cycles |
| --- | --- | --- |
| Crossbar.250 | 5.180200 | 2.03E+05 |
| Crossbar.125 | 6.4275373 | 2.65E+06 |
| 2d torus.250 | 3.5924281 | 8.2E+04 |
| 2d torus.125 | 4.8717425 | 1.07E+06 |
| Hypercube.250 | 3.7425749 | 9.8E+04 |
| Hypercube.125 | 5.2338263 | 9.98E+05 |

We see that as the bandwidth is increased, the link utilization decreases a little. But the average contention cycles decreases by an order. Hence, there is less opportunity for CAM to prioritize critical request over non-critical request. This results in less speedup when the bandwidth is less constrained.

**3.4 Performance with Splash2 benchmarks**

As the performance of CAM is highly dependent on the number of critical section, we choose cholesky and water-spatial from Splash2 benchmark suite [3] which has some locks for our study.

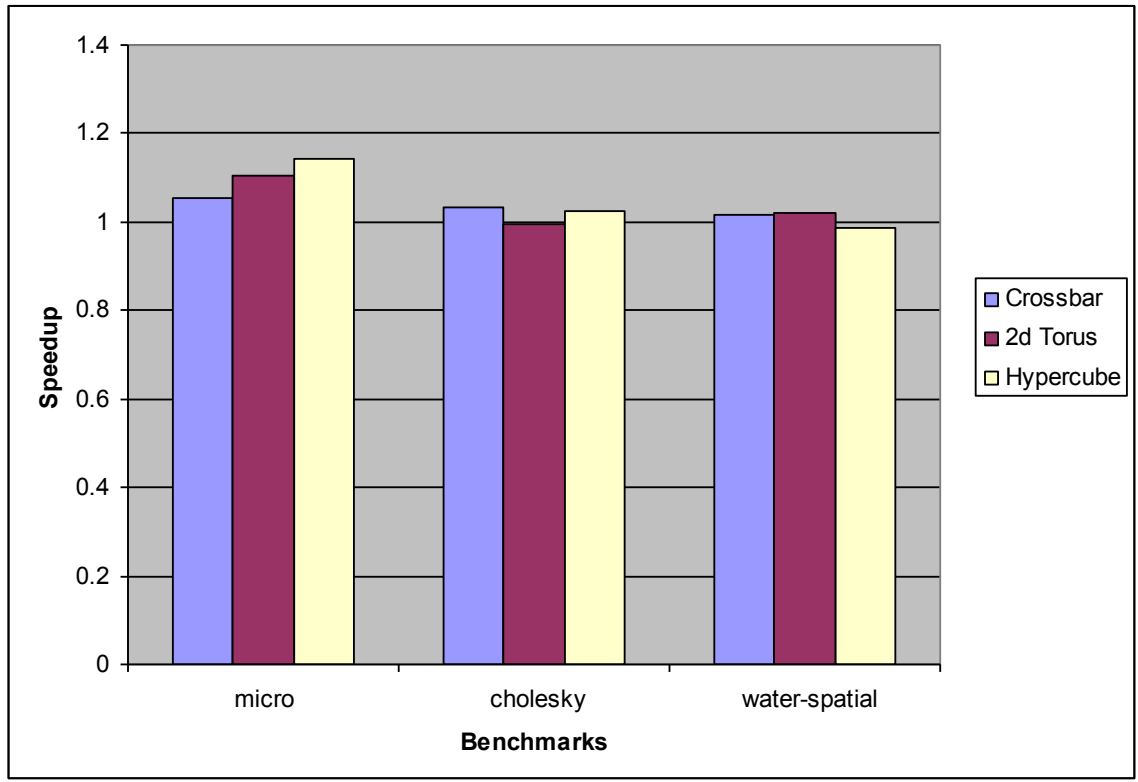

For a 16 processor system, we see that CAM gives a speedup of 1-3% for cholesky and almost negligible speedup for water-spatial.

To find out why cholesky gives little speedup and water-spatial gives almost no speedup, we again try to find out the ratio of critical and non-critical requests.

| Bench | crit-req | non-crit-req | Ratio |
| --- | --- | --- | --- |
| micro.300.16p.CROSSBAR.0125 | 298038 | 479900 | 0.383113 |
| Cholesky.16p.CROSSBAR.0125 | 41734 | 1063488 | 0.0377607 |
| Water-spatial.16p.CROSSBAR.0125 | 380 | 485056 | 0.000782801 |

We see that cholesky has some critical requests. But water-spatial does not have any critical requests.

## 4. Conclusions and Future work:

Criticality aware multiprocessor provides a new direction for tapping performance in a shared memory multiprocessor and can provide substantial speedup in lock intensive benchmarks.

To tap the full potential of criticality aware multiprocessor, there are many avenues which needs further investigation.

1) Benchmarks: It would be interesting to see how CAM performs in a real benchmark suite which has lot of locks.
2) Switches: A more realistic modeling of switch could result in more contention between critical and non-critical requests. This could result in CAM performing even better.
3) Barriers: Identifying other sources of criticality like barriers can help increasing the performance of criticality aware multiprocessor
4) Out of order processor model and CMPs[6]: Present investigation used a simple inorder model and SMP. It would be interesting to see how CAM performs using an out of order processor model/CMP.